\documentclass[twocolumn,preprintnumbers,amsmath,amssymb,superscriptaddress]{revtex4}
\usepackage[utf8x]{inputenc} 

\usepackage{graphicx}
\usepackage{dcolumn}
\usepackage{bm}
\usepackage{soul}
\usepackage{color}
\usepackage{epstopdf}
\usepackage[version=3]{mhchem}
\usepackage{lipsum}
\usepackage[outercaption]{sidecap}
\usepackage{floatrow}
\begin{document}

\title{Impact of high pressure on reversible structural relaxation of metallic glass}

\author{Nguyen K. Ngan}
\affiliation{Faculty of Materials Science and Engineering, Phenikaa University, Hanoi 12116, Vietnam}
\email{ngan.nguyenkim@phenikaa-uni.edu.vn}

\author{Anh D. Phan}
\affiliation{Faculty of Materials Science and Engineering, Phenikaa University, Hanoi 12116, Vietnam}
\affiliation{Phenikaa Institute for Advanced Study, Phenikaa University, Hanoi 12116, Vietnam}
\email{anh.phanduc@phenikaa-uni.edu.vn}

\author{Alessio Zaccone}
\affiliation{Cavendish Laboratory, University of Cambridge, JJ Thomson Avenue, CB3 0HE Cambridge, United Kingdom}
\affiliation{Statistical Physics Group, Department of Chemical Engineering and Biotechnology, University of Cambridge, Philippa Fawcett Drive, CB3 0AS Cambridge, United Kingdom}
\affiliation{Department of Physics "A. Pontremoli", University of Milan, via Celoria 16, 20133 Milano, Italy}
.

\date{\today}

\begin{abstract}
We theoretically investigate the temperature dependence of the reversible structural relaxation time and diffusion constant of metallic glasses under pressure. The compression not only changes the glassy dynamics, but also generates a metastable state along with a higher-energy state where the system can rejuvenate. The relaxation times for forward and backward transitions in this two-state system are nearly identical and much faster than the relaxation time without accounting for barrier-recrossing. At ambient pressure, the expected irreversible relaxation process is recovered, and our numerical results agree well with prior experimental results. An increase of pressure has a minor effect on the relaxation time and diffusion constant that one computes without considering the influence of the metastable state, but it leads to a large reduction of the reversible relaxation time computed upon taking the metastable state into account. The presence of external compression is also shown to trigger a fragile-to-strong crossover in metallic glasses. 
\end{abstract}

\maketitle


Metallic glasses are innovative metal alloy materials with disordered atomic structures, which lead to various advantages compared to their crystalline counterparts. To obtain amorphous structures, metallic glasses are cooled down very rapidly upon heating to avoid crystallization \cite{Deb2001}. Inheriting significant characteristics from both metals and plastics, amorphous metals exhibit exceptional mechanical properties and performance, including high strength, resilience, corrosion resistance and great flexibility but light weight, with promising applications in military, engineering and even and sport materials  \cite{Sun2015, Schuh2007, Inoue2000, Sun2018, Huang2017}. 
These properties strongly depend on structural relaxation process in metallic glasses. Therefore, to improve such properties, it is crucial to understand the glass transition and the underlying atomistic dynamics. 

There are several approaches to investigate glassy dynamics of metallic glass. Experimental scientists determine the pressure and temperature dependence of the relaxation process and the mechanical properties by using inelastic scattering, dynamic mechanical analysis (DMA) or mechanical spectroscopy, differential scanning calorimetry, and thermomechanical analysis. Molecular dynamics (MD) simulations have been employed to provide systematic calculations, understand time evolution of atomic positions and how components interact with each other in alloys and composites. Nevertheless, the timescale of simulations is on the order of many picoseconds, which is orders of magnitude larger than that accessible in experiments of supercooled and glassy systems ($\sim 100$ s).
Recently, the Elastically Collective Nonlinear Langevin Equation (ECNLE) theory was introduced and has been developed to study structural relaxation time and dynamic fragility of various materials including amorphous drugs, metallic glasses, and colloidal systems \cite{Phan2018, Phan2019, PhanPRL} under different compression protocols. ECNLE calculations have quantitatively and qualitatively described experimental results.

In most amorphous materials, the structural relaxation is an irreversible process or, in other words, the energy profile has a single basin or (inherent) state from which the system relaxes away. Interestingly, a secondary state could arise with external energy injection and it may reverse the structural relaxation. Previous works showed that the two-state level system or structural rejuvenation can be achieved by thermal cycling \cite{Ketov2015}, thermomechanical creep \cite{Tong2018}, elastostatic loading \cite{Park2008}, shock compression \cite{Ding2019}. However, the physical mechanism underlying this phenomenon has remained poorly understood. 

In this Letter, we propose a theoretical approach based on the ECNLE theory to calculate the temperature dependence of structural relaxation time and diffusion constant of metallic glasses under external pressures. We consider amorphous materials as a dense system of discrete spherical particles interacting with each other via hard-sphere interaction. The key parameters characterizing the theoretical system include the particle diameter, $d$, and the number of particles per volume, $\rho$. Thus, the volume fraction is $\phi = \rho\pi d^3/6$. Under ambient pressure, the dynamic free energy of a tagged particle caused by the nearest neighbor constraint is \cite{Phan2018,3,4,6}

\begin{eqnarray}
\label{eq:2}
\frac{F_{dyn}(r)}{k_BT} &=& -3\ln\frac{r}{d}
\\ &-&\int_0^{\infty} \textrm{d}q\frac{ q^2d^3 \left[S(q)-1\right]^2}{12\pi\phi\left[1+S(q)\right]}\exp\left[-\frac{q^2r^2(S(q)+1)}{6S(q)}\right],\nonumber
\end{eqnarray}
where $r$ is the scalar displacement of the tagged particle, $k_B$ is the Boltzmann constant, $T$ is temperature, $q$ is the wavevector, and $S(q)$ is the static structure factor. $S(q)$ can be calculated using MD simulations \cite{Allen2004}, Percus-Yevick theory \cite{1}, or polymer reference interaction site model (PRISM) theory \cite{pyPrism2018}. In our work, we use PRISM theory. From this, we obtain the radial distribution function, $g(r)$, extracted using the Fourier transform of the structure factor, $g(r)=1+\cfrac{1}{2\pi^2\rho r} \int_0^{\infty}\left[S(q)-1\right]q\sin(qr)dq$. The first term in rhs of Eq.~(\ref{eq:2}) corresponds to the ideal fluid state. Meanwhile, the second term contains information of structure and density components to describe the caged dynamics leading to particle localization.

In the presence of external pressure $P$, the dynamic free energy becomes \cite{PhanPRL,PhanPCCP,PhanPhar,PhanACSO} 

\begin{eqnarray}
\frac{F_{dyn}(r,P)}{k_BT}=\frac{F_{dyn}(r)}{k_BT} + \frac{P}{k_BT/d^3}\frac{r}{d}.
\label{eq:Fdyn_withP}
\end{eqnarray}

\begin{figure}[htp]
\center
\includegraphics[width=8cm]{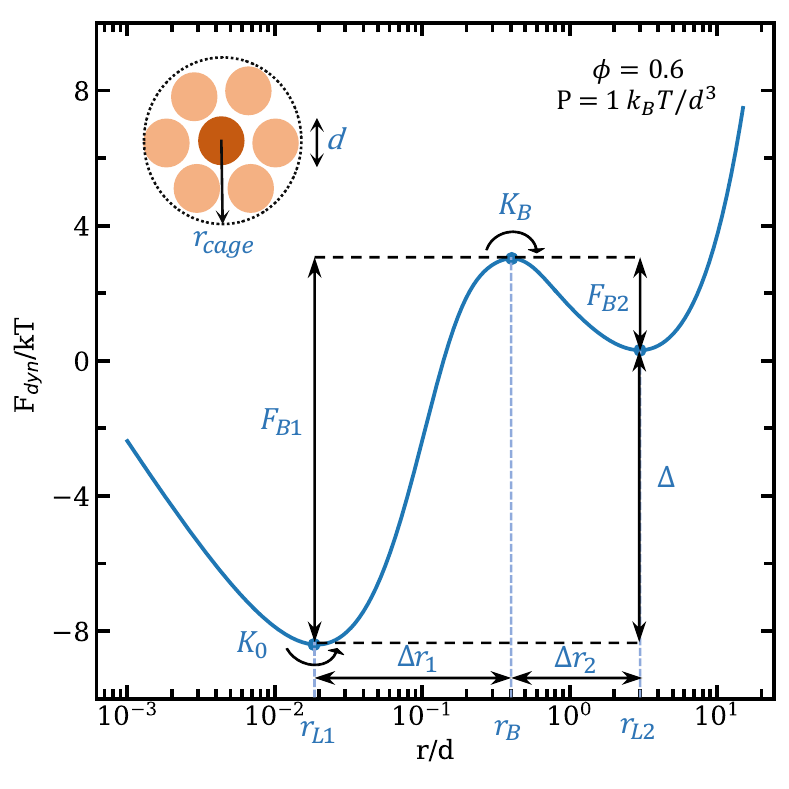}
\caption{(Color online) The dynamic free energy of a tagged particle in a hard-sphere fluid of $\Phi=0.60$ under compression $P=1k_BT/d^3$. Key length scales and two barriers in the two-state level system are indicated.}
\label{fig1:illustration}
\end{figure}

Figure~\ref{fig1:illustration} shows an example of the profile of the dynamic free energy when an external pressure is applied. At sufficiently high density ($\Phi > 0.43$) \cite{3,4,1}, the free volume is significantly reduced, and the tagged particle is confined within a particle cage formed by its nearest neighbors. Thus, a free-energy barrier emerges. The particle cage radius, $r_{cage}$, illustrated in Figure~\ref{fig1:illustration} is estimated as the first minimum position of $g(r)$. $F_{dyn}(r)$ or $F_{dyn}(r, P = 0)$ in Equation~(\ref{eq:Fdyn_withP}) is mainly responsible for the primary (first) inherent state characterized by the localization length, $r_{L1}$, the barrier position, $r_B$, and the barrier height, $F_{B1}=F_{dyn}(r_B, P)-F_{dyn}(r_{L1}, P)$. In the absence of external pressure, the secondary state does not occur and the relaxation is an irreversible process. Under compression, a metastable state emerges and the relaxation becomes reversible. One can see another minimum position, $r_{L2}$, and barrier height, $F_{B2}=F_{dyn}(r_B,P)-F_{dyn}(r_{L2},P)$.  At this second minimum, the effect of applied pressure dominates the caging restriction \cite{PhanPRL, PhanACSO}. The distances between the barrier position and these two minimum locations are jump distances calculated as $\Delta r_1 =r_B-r_{L1}$ and $\Delta r_2 =r_{L2}-r_B$.

For metallic glasses, without considering effects of the secondary state, the $\alpha$ relaxation time for the tagged particle to escape across the primary (first) barrier (left to right) in the dynamic free energy (as shown in Figure~\ref{fig1:illustration}) is calculated using Kramers' escape rate theory

\begin{eqnarray}
\frac{\tau_\alpha}{\tau_s} = 1+ \frac{2\pi}{\sqrt{K_0K_B}}\frac{k_BT}{d^2} e^{F_{B1}/k_BT},
\label{eq:6}
\end{eqnarray}
where $K_0 = \left|\partial^2 F_{dyn}(r,P)/\partial r^2\right|_{r=r_{L1}}$ and $K_B$=$\left|\partial^2 F_{dyn}(r,P)/\partial r^2\right|_{r=r_B}$ are absolute curvatures at $r_{L1}$ and $r_B$, respectively, and $\tau_s$ is a short relaxation time scale. Effects of collective elastic motions on the structural relaxation of metallic glasses are very small \cite{PhanPRL}. Thus, we also ignore it in our calculations. 

The analytical expression for $\tau_s$ is \cite{6,7}
\begin{eqnarray}
\tau_s=g^2(d)\tau_E\left[1+\frac{1}{36\pi\Phi}\int_0^{\infty}\textrm{d}q\frac{q^2(S(q)-1)^2}{S(q)+b(q)} \right],
\label{eq:taus}
\end{eqnarray}
where $b(q)=1/\left[1-j_0(q)+2j_2(q)\right]$, $j_n(x)$ is the spherical Bessel function of order $n$, and $\tau_E$ is the Enskog time scale. In prior works \cite{Phan2019, PhanPRL}, $\tau_E =$ 0.1 ps was found for metallic glasses.

The above structural relaxation time can be interpreted as a forward hopping time in an irreversible process. The fact that the presence of the secondary state leads to reversibility of the structural relaxation or backward hopping thus alters $\tau_\alpha$.

Here we propose a simple approximation based on the two-level system model to determine the temperature dependence of the pressure-induced reversible relaxation process, which takes into account effects of the recrossing phenomenon on glassy dynamics. Let us call $p_1(t)$ and $p_2(t)$ the probabilities to find the tagged particle in the state 1 and 2, respectively, then we have
\begin{eqnarray}
dp_1(t) = -p_1(t)e^{-\frac{F_{B1}}{k_BT}}\nu_0dt+p_2(t)e^{-\frac{F_{B2}}{k_BT}}\nu_0dt,
\label{eq:prob1}
\end{eqnarray}
where $\nu_0$ is the attempt frequency. The first and second term on the r.h.s. of Eq.~(\ref{eq:prob1}) correspond to the transition from state 1 to 2 and vice versa, respectively. Since the particle surely exists in one of two states, $p_1(t) + p_2(t) = 1$. We suppose that the particle is located in the state 1 at $t = 0$, then an analytical expression for $p_1(t)$ is
\begin{eqnarray}
p_1(t) = \frac{e^{-\frac{F_{B2}}{k_BT}} + e^{-\frac{F_{B1}}{k_BT}}e^{-\nu_0t[e^{-F_{B1}/k_BT}+e^{-F_{B2}/k_BT}]} }{e^{-\frac{F_{B1}}{k_BT}}+e^{-\frac{F_{B2}}{k_BT}}},
\label{eq:prob2}
\end{eqnarray}

The reversible relaxation time is determined by the time at which the time-dependent component of $p_1(t)$ is reduced by a factor of $e$. Thus, this $\alpha$ time for the forward transition, from state 1 to 2, is given by
\begin{eqnarray}
\tau_{\alpha,12}^*&=& \frac{1}{\nu_0[e^{-F_{B1}/k_BT}+e^{-F_{B2}/k_BT}]} \nonumber\\
&=& \frac{\nu_0^{-1}e^{F_{B1}/k_BT}}{1+e^{(F_{B1}-F_{B2})/k_BT}} = \frac{\tau_{\alpha,12}}{1+e^{\frac{\Delta}{k_BT}}},
\label{eq:tau*}
\end{eqnarray}
where $\Delta=F_{B1}-F_{B2}$ is the energy difference between two states in the energy profile of compressed systems (see Figure~\ref{fig1:illustration}) and $\tau_{\alpha,12}=\nu_0^{-1}e^{F_{B1}/k_BT}$ according to the Vogel-Fulcher-Tammann (VFT) function \cite{VFTfunction}. Equation~(\ref{eq:tau*}) suggests that an increase of $\Delta$ diminishes the forward hopping time. Under ambient pressure condition $(P = 0)$, there is no secondary state, $F_{B2}=\infty$ and $\Delta = -\infty$, we have $\tau_{\alpha,12}^*=\tau_{\alpha,12}$ and the reversible calculation reduces to the irreversible one. In the presence of external pressure, probabilities of the particle existing in the steady state 1 and 2 are, respectively,
\begin{eqnarray}
p_1(t=\infty)=\frac{e^{\frac{\Delta}{k_BT}}}{1+e^{\frac{\Delta}{k_BT}}}, \quad  p_2(t=\infty)=\frac{1}{1+e^{\frac{\Delta}{k_BT}}}.
\label{eq:p}
\end{eqnarray}
Thus, the reversible relaxation time can be calculated in an average manner as
\begin{eqnarray}
\tau_{\alpha,12}^* = p_2(t=\infty)\tau_{\alpha,12}= \frac{\tau_{\alpha,12}}{1+e^{\frac{\Delta}{k_BT}}}.
\label{eq:tau12*}
\end{eqnarray}

Similarly, the time scale for backward transition, from state 2 to 1, is
\begin{eqnarray}
\tau_{\alpha,21}^*&=& \frac{\nu_0^{-1}e^{F_{B2}/k_BT}}{1+e^{(F_{B2}-F_{B1})/k_BT}} \nonumber\\
&=& \frac{\nu_0^{-1}e^{F_{B1}/k_BT}}{1+e^{(F_{B1}-F_{B2})/k_BT}} 
=\tau_{\alpha,12}^*.
\label{eq:tau21*}
\end{eqnarray}

Equation~(\ref{eq:tau21*}) suggests that the forward and backward transitions happen on the same (constant) time scale $\tau_{\alpha,12}^* = \tau_{\alpha,21}^* = \tau_{\alpha}^*$. The difference between our $\tau_{\alpha}^*$ and a classical formula for the relaxation time proposed by Gilroy and Phillips \cite{Gilroy1981} is just a factor of 2. The reason for this difference is that they determined the back-and-forth relaxation time in an averaged manner, that is $\cfrac{\tau_{\alpha,12}^*\tau_{\alpha,21}^*}{\tau_{\alpha,12}^*+\tau_{\alpha,21}^*}=\cfrac{\tau_{\alpha}^*}{2}$. 





Based on thermal expansion, a density-to-temperature conversion (thermal mapping) is proposed \cite{PhanPRL,PhanPCCP,PhanPhar,PhanACSO,42} to determine the temperature dependence of the relaxation time. The thermal mapping is given by
\begin{eqnarray}
T \approx T_g + \frac{\phi_g-\phi}{\beta\phi_0},
\label{eq:7}
\end{eqnarray} 
where $T_g$ is the experimental glass transition temperature at ambient pressure, $\beta$ is an effective volume thermal expansion coefficient, and $\phi_0$ and $\phi_g$ are the characteristic volume fraction and the volume fraction at the glass transition ($\tau_\alpha(\phi_g)=100$ s), respectively. For metallic glasses and other amorphous materials \cite{PhanPRL,PhanPCCP,PhanPhar,PhanACSO,42}, $\phi_0 = 0.50$ and $\beta\phi_0 = 6\times 10^{-4}$ $\textrm{K}^{-1}$ \cite{42}, however, $\phi_g=0.6585$ for metallic glasses \cite{PhanPRL}. 



\begin{figure*}[htp]
\center
\includegraphics[width=15cm]{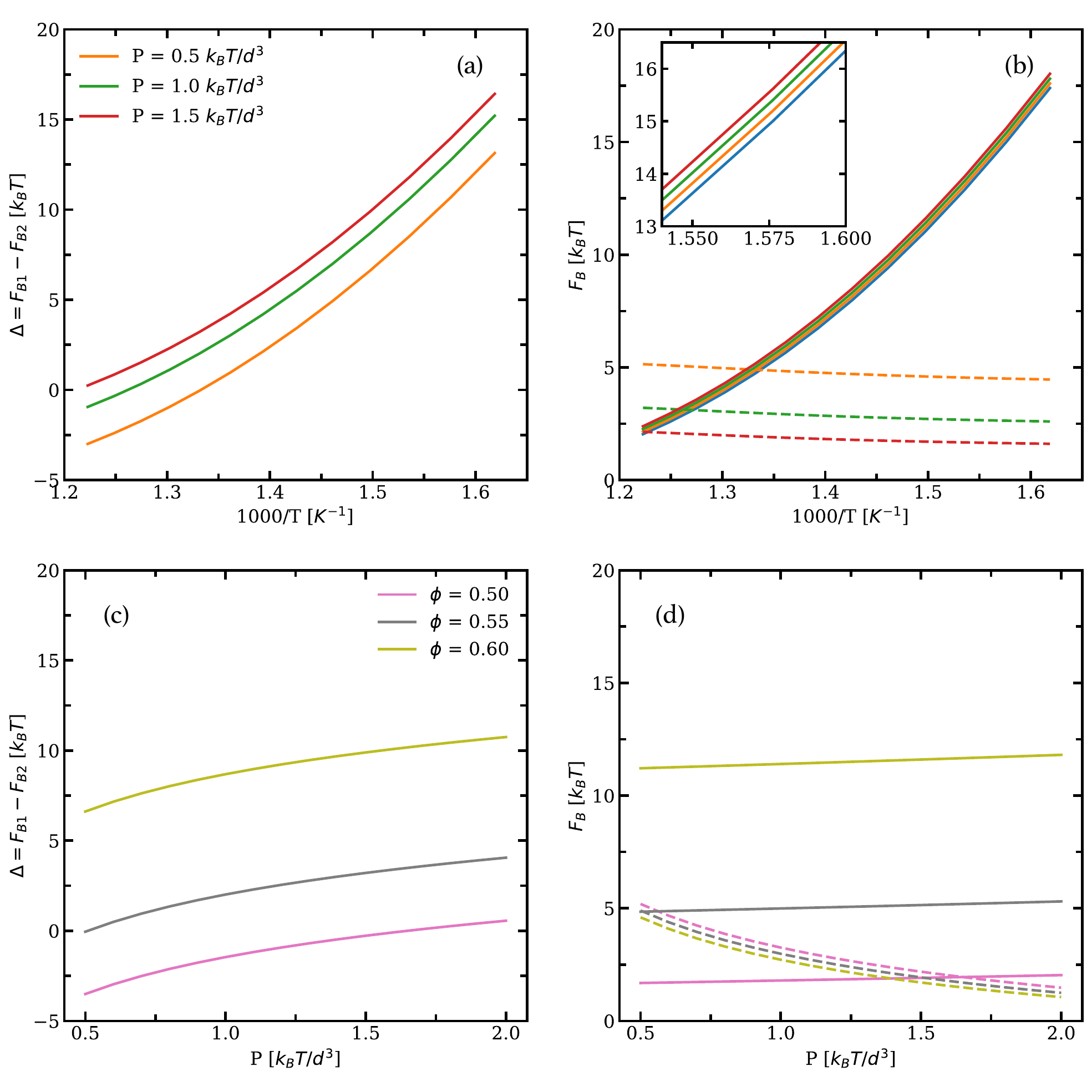}
\caption{(Color online) The temperature dependence of (a) $\Delta$ and (b) $F_B$ at $P/(k_BT/d^3) = 0.5$, 1.0, and 1.5. The inset zooms in on a small region of data of the main frame. The blue line indicates the free external pressure case. Panels (c) and (d) show the pressure dependencies of $\Delta$ and $F_B$, respectively, calculated at $\phi$ = 0.50, 0.55 and 0.60. The solid and dashed curves correspond to the primary and secondary local hopping barriers ($F_{B1}$ and $F_{B2}$ calculated in the text), respectively.}
\label{fig2:FB}
\end{figure*}

Figure~\ref{fig2:FB} shows the temperature and pressure dependence of $F_{B1}$, $F_{B2}$, and $\Delta=F_{B1}-F_{B2}$. At low temperatures, the density of materials is high and it leads to a reduction of the free volume. Thus, the caging constraints are strengthened and $F_{B1}$ grows with cooling as shown in Figure~\ref{fig2:FB}b. Meanwhile, previous work \cite{PhanPRL} indicated that $r_{L2} = 3k_BT/Pd^2$ is rather insensitive to temperature, and the role of the nearest-neighbor interactions can be ignored. This explains why lowering temperature leads to a slight decrease of $F_{B2}$ and an increase of $\Delta$ at a given pressure in Figure~\ref{fig2:FB}a.

At a given temperature, the energy difference between two localized states $\Delta$ increases with pressure since $F_{B1}$ slightly increases but $F_{B2}$ is significantly reduced as shown in Figure~\ref{fig2:FB}c and \ref{fig2:FB}d. When the external pressure is larger than a threshold value, the secondary local minimum disappears and $F_{B2}=0$. Then the energy landscape has a potential well with a virtually infinite barrier \cite{PhanPRL}. The relaxation time goes to infinity and the tagged particle seems to be trapped in a pseudo-crystalline state and cannot escape from its particle cage unless the interatomic repulsions are overcome.



\begin{figure*}[htp]
\center
\includegraphics[width=15cm]{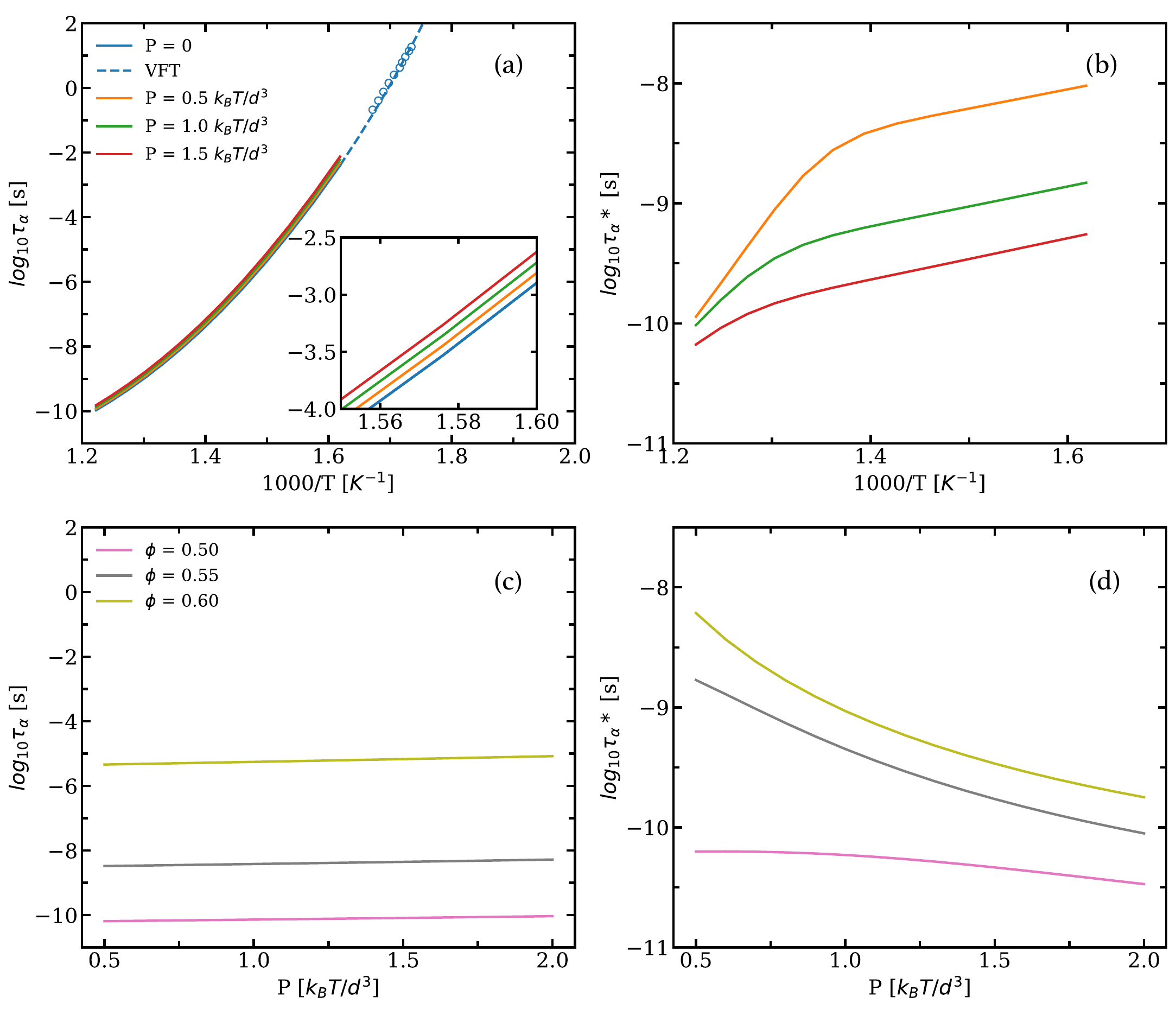}
\caption{(Color online) The temperature dependence of (a) log$_{10}\tau_{\alpha}$ (Equation~(\ref{eq:6})) and (b) log$_{10}\tau_{\alpha}^*$ (Equation~(\ref{eq:tau12*})) at $P/(k_BT/d^3) = 0, 0.5$, 1.0, and 1.5. Open points correspond to experimental data of $\ce{Pd_{40}Ni_{10}Cu_{30}P_{20}}$ taken from Ref. \cite{Qiao2014}. Blue dashed line indicates an extension of the theoretical curve (P = 0) using the VFT equation: ln$\tau_{\alpha}=-44.65+9062.25/(T-386.25)$. The inset zooms in on a small region of data of the main frame. Panels (c) and (d) show the pressure dependence of log$_{10}\tau_{\alpha}$ and log$_{10}\tau_{\alpha}^*$, respectively, at $\phi$ = 0.5, 0.55 and 0.6.}
\label{fig3:tau}
\end{figure*}

Figure~\ref{fig3:tau} shows how reversible and irreversible relaxation times depend on temperature and pressure. We theoretically calculate $\log\tau_\alpha$ and $\log\tau_\alpha^*$ as a function of $1000/T$ by using Equation~(\ref{eq:6}), (\ref{eq:taus}), and (\ref{eq:tau12*}) at pressures $P/(k_BT/d^3)$ = 0, 0.5, 1.0, 1.5. Numerical results from theory and experimental data are compared in Figure \ref{fig3:tau}a. Since $\tau_\alpha$ and $\tau_\alpha^*$ are proportional to $e^{F_{B1}/k_BT}$, both the forward and backward relaxation times vary with temperature in the same manner as $F_{B1}$. The dynamics is significantly slowed down at low temperatures. The presence of external pressure has a minor effect on $\tau_{\alpha}(T)$. This finding is consistent with experimental data for glycerol, propylene carbonate, and metafluoroaniline when the external pressure is increased up to 700 MPa in Ref. \cite{Reiser2006}.

Intrinsic limitations of the pyPRISM code restrict the timescale accessible in our calculation. As $S(q)$ and $g(r)$ cannot be computed beyond $\phi=0.64$, our relaxation time only ranges from 100 ps to 0.01 s. Thus, we extrapolate the theoretical curve at $P = 0$ (as an example) using the Vogel-Fulcher-Tamman (VFT) function to compare to experimental results \cite{Qiao2014}. The VFT form for structural relaxation time is \cite{VFTfunction}
\begin{equation}
\textrm{ln}\tau = \textrm{ln}\tau_0 + \frac{B}{T-T_0},
\label{eq:VFT}
\end{equation}
where the fitting parameters are ln$\tau_0 = -44.65 \pm 0.6$, $B = 9062.25 \pm 314.59$, $T_0 = 386.25 \pm 4.73$. The VFT calculation is shown as the blue dashed curve in Figure~\ref{fig3:tau}a and it shows a quantitative agreement between the ECNLE theory and the experimental results.



Equation~(\ref{eq:tau12*}) suggests that the reversible hopping time $\tau_{\alpha}^*$ is much less than the irreversible counterpart $\tau_{\alpha}$. The difference between the two relaxation times is a factor of ($1+e^{\Delta/k_BT}$). One can directly compare numerical results via Figure~\ref{fig3:tau}a and \ref{fig3:tau}b. Since the compression induces an increase of $\Delta$, we obtain a decrease of $\tau_{\alpha}^*$ with pressure in Figure \ref{fig3:tau}b and \ref{fig3:tau}d. This trend is opposite to the pressure-induced variation of $\tau_{\alpha}$. The weak pressure dependence of $F_{B1}$ at a given density or temperature leads to a very small increase of $\log\tau_\alpha$ with increasing $P$ as shown in Figure~\ref{fig3:tau}c. However, the pressure effect diminishes moderately in the system having low density or large free volume. 





The diffusion constant $D$ describing how much the particle diffuses in the glassy environment is calculated via

\begin{eqnarray}
D =\frac{\Delta r_1^2}{6\tau_{\alpha}}.
\label{eq:D}
\end{eqnarray}

\begin{figure*}[htp]
\center
\includegraphics[width=\textwidth]{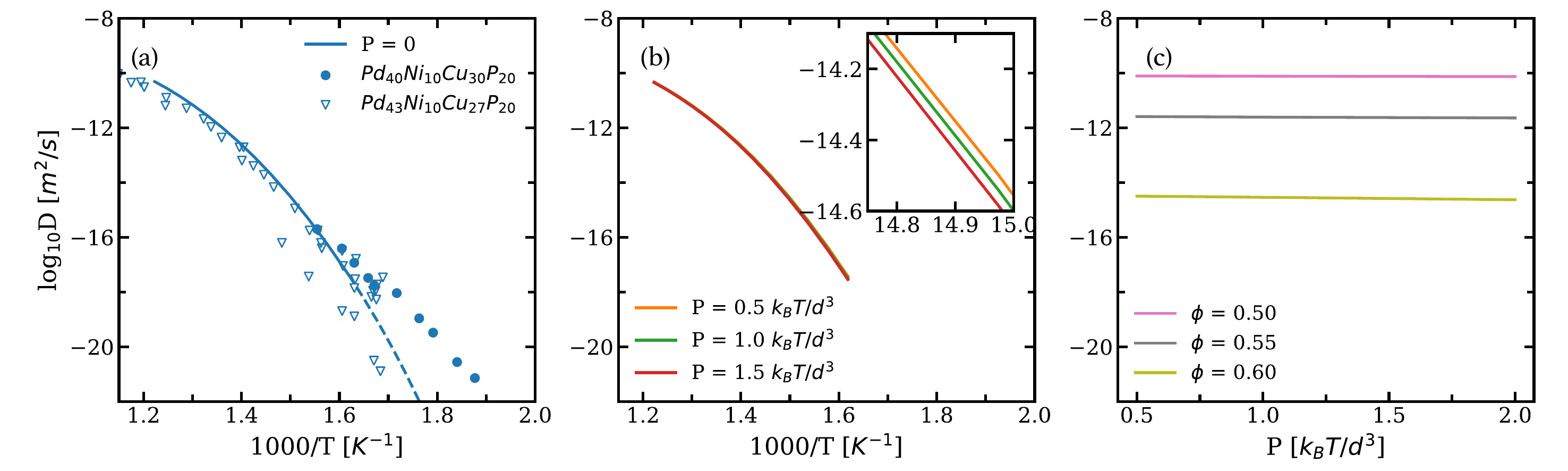}
\caption{(Color online) The dependence of log$_{10}$D on temperature at ambient conditions (a), 0.5, 1.0, 1.5 (b); and on pressure (c) at $\phi$ = 0.5, 0.55 and 0.6. Open and solid data points are experimental results of $\ce{Pd_{43}Ni_{10}Cu_{27}P_{20}}$ and $\ce{Pd_{40}Ni_{10}Cu_{30}P_{20}}$, respectively, taken from References~\cite{Bartsch2010,Kia2017}. Blue solid and dashed curves correspond to ECNLE and VFT calculations based on the predicted $\tau_\alpha$ in Figure~\ref{fig3:tau}. The inset zooms in on a small region of data of the main frame.}
\label{fig4:diffusion const}
\end{figure*}

Figure~\ref{fig4:diffusion const}a shows the theoretical and experimental temperature dependence of diffusion constant for $\ce{Pd_{40}Ni_{10}Cu_{30}P_{20}}$ and $\ce{Pd_{43}Ni_{10}Cu_{27}P_{20}}$ at ambient pressure. Since previous works \cite{Phan2018,6,PhanACSO} used a particle diameter $d$ for glassy materials in range of 0.4 - 1.2\,nm, we chose $d=0.8$\,nm for our calculations of $D$. Although experimental data are noisy, our prediction is quantitatively close to experiments. Our calculation is consistent with a larger diffusion at higher temperatures. Note that Equation~(\ref{eq:D}) is applicable to other amorphous materials including drugs and polymers. In the case of amorphous drugs, this model can be exploited to determine thermal-induced enhancement of drug solubility. Since the compression slows down the dynamics by a less than an order of magnitude, only a minor enhancement of the diffusion constant is expected when applying pressure as one can clearly see from numerical results in Figure~\ref{fig4:diffusion const}b and \ref{fig4:diffusion const}c. At a certain temperature, $D(P)$ of metallic glasses remains nearly unchanged.

The dynamic fragility is characterized by the dimensionless steepness index:
\begin{eqnarray}
m=\left. \frac{\textrm{dlog}_{10}\tau_\alpha}{\textrm{d}(T_g/T)}\right|_{T=T_g}.
\label{eq:m}
\end{eqnarray}

The glass is denominated ``strong'' if $m \leq 30$, and "fragile'' if $m \geq 100$. By fitting the $\tau_\alpha$ with the VFT function (Eq.~(\ref{eq:VFT})), the fragility index can be obtained via the relationship:
\begin{eqnarray}
m=\frac{BT_g}{\textrm{ln}10(T_g-T_0)^2}.
\label{eq:m with VFT}
\end{eqnarray}

\begin{figure}[htp]
\center
\includegraphics[width=8cm]{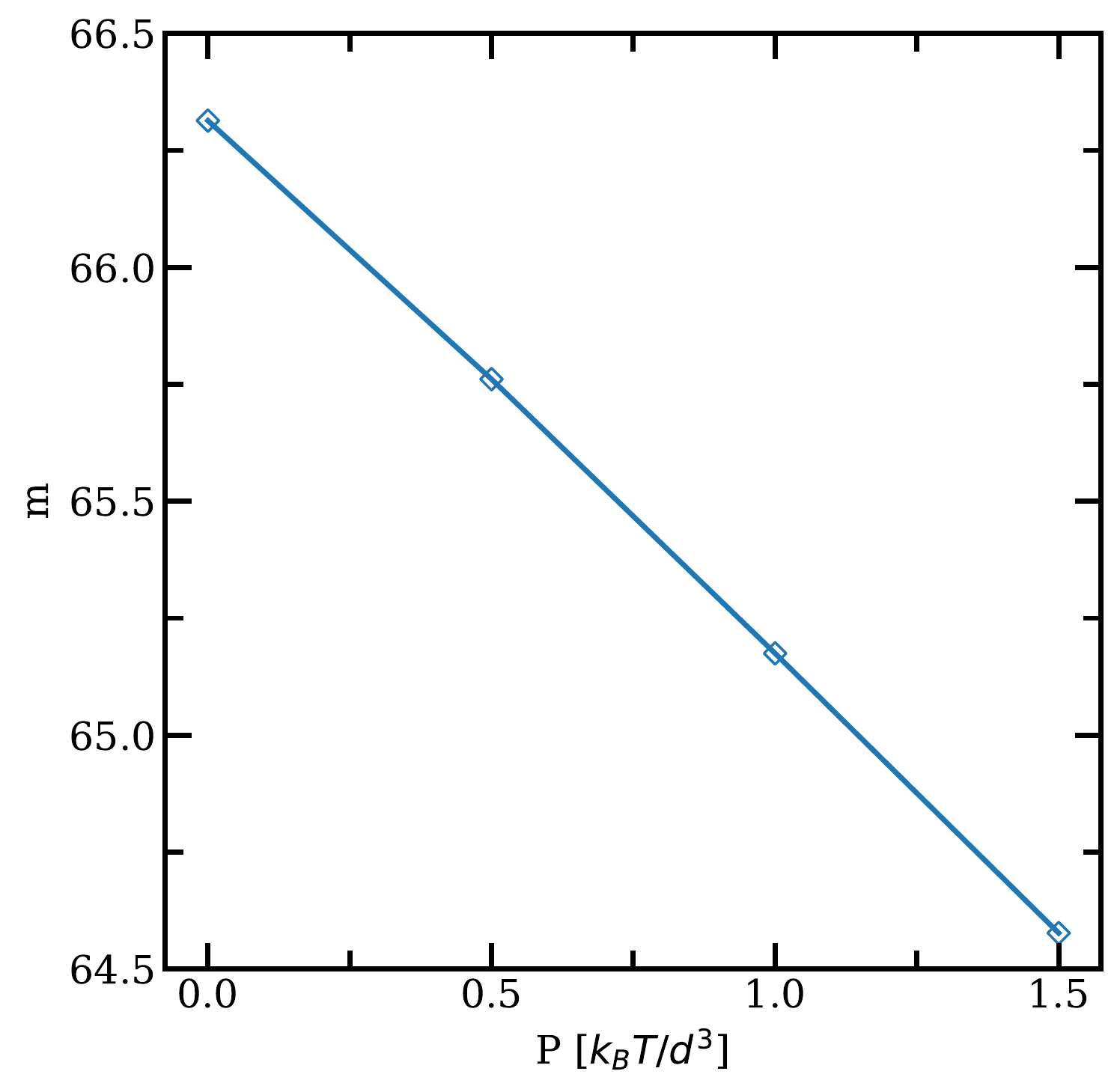}
\caption{(Color online) The dependence of fragility on pressure of $\ce{Pd_{40}Ni_{10}Cu_{30}P_{20}}$ under compression.}
\label{fig5:m}
\end{figure}

Figure~\ref{fig5:m} shows how the dynamic fragility of metallic glass changes under external pressure. The metallic glasses used in this simulation is neither strong nor fragile, but its fragility index diminishes with increasing pressure, thus presenting a fragile-to-strong crossover. Similar behaviors were experimentally reported for glass-forming liquids in \cite{Casalini2005, Roland2005}, while some simulations suggested a pressure-induced increase of the fragility of glassy materials \cite{Shintani2008, Hu2017}. Over the past decades, the relation between fragility and pressure has remained controversial. The challenging in this issue is reflected in Reference \cite{Reiser2006}, where authors found the the steepness index of different organic glass-forming liquids showed various behaviors (unchanged, increasing and decreasing) with increasing pressure.  

In small-molecule and polymer glass formers, effects of collective motions of molecules on the glassy dynamics and fragility \cite{Phan2019,Phan2018,6} are very important. The ECNLE theory quantifies these collective effects via the elastic barrier, and this barrier is coupled with the local barrier $F_{B1}$ in Eq. (\ref{eq:6}) to significantly increase the structural relaxation time. Thus, $\tau_\alpha=100$ s of organic materials occurs at smaller densities (or volume fractions) than densities of metallic glasses. This explains why $T_g$ of metallic glasses is much larger than that of small-molecule and polymer glasses. Furthermore, correlation between fragility and cooperativity in Ref. \cite{Phan2019} revealed that stronger cooperative rearrangement leads to larger fragility of materials, in agreement with models based on interatomic potential steepness~\cite{Krausser}. This is also consistent with the fact that organic materials are typically more fragile than metallic glasses. Under compression, both the local (potential) and the long-range (elastic) barriers become larger (whereas in metallic glass the variation of elastic barrier is negligible~\cite{PhanPRL}). In organic glass-forming liquids, the increase of both barriers causes slower molecular dynamics in comparison with the situation at ambient pressure at a given $\phi$, and thus a stronger pressure-dependence of structural relaxation and fragility. The ENCLE theory predicts the fragile-to-strong transition in metallic glasses but the process may be reversed in other materials having non-zero long-range elastic effects \cite{PhanJPCB,PhanACSO}.

We have calculated the temperature dependence of the irreversible and reversible relaxation time, and diffusion constant of a metallic glass under compression using the ECNLE theory. When considering the system at ambient pressure ($P \approx 0$), the system has a single localized state and the relaxation is irreversible. Numerical results for this case agree quantitatively with previous experimental data \cite{Qiao2014,Bartsch2010,Kia2017}. The presence of external pressure induces an emergence of an additional higher-energy secondary state which can reverse the relaxation process. The forward and backward relaxation times are equal and much shorter than the irreversible relaxation time. The compression has minor influence on the temperature dependence of the primary local barrier, irreversible relaxation time and diffusion constant. The backward barrier from the secondary state back to the primary low-energy state exists at low to moderate pressures and decreases with further increasing pressure. Thus, the energy difference between the two states varies greatly and causes the substantial variation of the reversible transition time. Although the dependence of fragility on external factors such as pressure is still a matter of debate, our prediction of a fragile-to-strong crossover due to an asymmetric double-well potential landscape arising under compression is consistent with previous works \cite{Casalini2005, Roland2005}. Since our approach has successfully described the glass transition of glass-forming liquids having simple and complex structures in both experiments \cite{Phan2018,3,4,6, PhanPRL,Phan2019,42,PhanJPCB} and simulation \cite{PhanACSO} in the past, in future work it will be possible to extend our theoretical framework to study more complex polymer glasses such as those in Ref. \cite{Wenjie2020-1,Wenjie2020-2}.


\end{document}